\documentclass[aps,prl,amsmath,amssymb,amsfonts,superscriptaddress,reprint]{revtex4-2}

\usepackage{dcolumn}%
\usepackage{siunitx}

\usepackage[normalem]{ulem}
\usepackage{bm}
\usepackage{amssymb}
\usepackage{amsfonts}
\usepackage{amsmath}

\usepackage{graphicx}

\usepackage{xcolor}

\usepackage[pdftex,colorlinks=true,allcolors=blue,breaklinks=true]{hyperref}

\usepackage[utf8]{inputenc}
\usepackage{textgreek}

\definecolor{g-blue}{rgb}{0.83,0.95,1}
\definecolor{g-yellow}{rgb}{1,1,0.7}
\definecolor{g-green}{rgb}{0.9,1,0.9}
\definecolor{green}{rgb}{0,0.6,0}
\definecolor{cyan}{rgb}{0,0.7,0.7}
\definecolor{black}{rgb}{0,0,0}
\definecolor{grey}{rgb}{0.4,0.4,0.4}
\definecolor{nature-blue}{rgb}{0.0,0.200,0.500}

\def \ed {\end{document}}
\def\Fbox#1{\vskip1ex\hbox to 8.5cm{\hfil\fboxsep0.3cm\fbox{%
		\parbox{8.0cm}{#1}}\hfil}\vskip1ex\noindent}

\def\be{\begin{equation}}
\def\ee{\end{equation}}
\def\bea{\begin{eqnarray}}
\def\eea{\end{eqnarray}}
\def\bse{\begin{subequations}}
\def\ese{\end{subequations}}

\let \= \equiv
\let\*\cdot
\let\^\widehat
\let\-\overline

\def\1{\bm1}

\def\<{\left\langle}    \def\>{\right\rangle}
\def\({\left(}          \def\){\right)}
\def\[ {\left[}         \def\]{\right]}

\newcommand{\RPTU}{\affiliation{Fachbereich Physik and Landesforschungszentrum OPTIMAS, Rheinland-Pf\"alzische Technische Universit\"at Kaiserslautern-Landau, 67663 Kaiserslautern, Germany}}
\newcommand{\UniWien}{\affiliation{University of Vienna, Faculty of Physics, 1090 Vienna, Austria}}
\newcommand{\UniAustralia}{\affiliation{University of Western Australia, Department of Physics M013, Crawley, 6009 WA, Australia}}

\newcommand{\Pthr}{P_\mathrm{thr}}
\newcommand{\Pp}{P_\mathrm{p}}
\newcommand{\fr}{f_\mathrm{r}}
\newcommand{\YIG}{\text{YIG}}
\newcommand{\GGG}{\text{GGG}}

\newcommand{\Figref}[1]{Fig.\,\ref{#1}}

\makeatletter
\AtBeginDocument{\let\SM@hangfrom@section\@hangfrom@section}
\newcommand{\SupplementalMaterialSetup}{%
  \onecolumngrid
  \clearpage
  \setcounter{secnumdepth}{-10}%
  \let\@hangfrom@section\SM@hangfrom@section
  \let\@sectioncntformat\relax
  \@removefromreset{equation}{section}%
  \setcounter{section}{0}\setcounter{subsection}{0}\setcounter{subsubsection}{0}%
  \setcounter{equation}{0}\setcounter{figure}{0}\setcounter{table}{0}%
  \renewcommand{\thesection}{S\Roman{section}}%
  \renewcommand{\theequation}{S\arabic{equation}}%
  \renewcommand{\thefigure}{S\arabic{figure}}%
  \renewcommand{\thetable}{S\arabic{table}}%
  \renewcommand{\theHsection}{SM.\arabic{section}}%
  \renewcommand{\theHequation}{SM.\arabic{equation}}%
  \renewcommand{\theHfigure}{SM.\arabic{figure}}%
  \renewcommand{\theHtable}{SM.\arabic{table}}%
}
\makeatother
\hypersetup{hypertexnames=false}

\begin{document}
\title[Physical Review Letters]
{Anisotropic wavevector-dependent damping of thickness-quantized magnons}

    \author{Tamara~Azevedo}
    \RPTU

    \author{Rostyslav~O.~Serha}
    \email{rostyslav.serha@univie.ac.at}
    \UniWien

    \author{Yannik~Kunz}
    \RPTU

    \author{Matthias~R.~Schweizer}
    \RPTU

    \author{Vitaliy~I.~Vasyuchka}
    \RPTU

    \author{Mathias~Weiler}
	\RPTU

    \author{Andrii~V.~Chumak}
    \UniWien

    \author{Burkard~Hillebrands}
	\RPTU

    \author{Mikhail~Kostylev}
    \email{mikhail.kostylev@uwa.edu.au}
    \UniAustralia

	\author{Alexander~A.~Serga}
 	\email{serha@rptu.de}
	\RPTU

\date{\today}%

\begin{abstract}
Magnon damping is a key factor governing spin-wave transport and nonlinear dynamics of multimode magnon systems. However, many descriptions rely on the assumption of an effective mode-independent parameter, which can mask wavelength-dependent relaxation processes that depend on propagation geometry and mode profile. Here, we employ high-resolution parametric-instability spectroscopy to probe thickness-quantized spin waves with wavelengths down to about a hundred nanometers in micrometer-thick yttrium iron garnet films. The instability threshold exhibits a regular sawtooth dependence on the magnetic field, arising from switching between discrete thickness modes. Comparison with dipole--exchange theory reveals anisotropic wavevector-dependent damping that increases with mode number and depends differently on the in-plane and out-of-plane wavevector components.
\looseness=-1
\end{abstract}

\maketitle

Magnon damping governs coherent spin-wave transport, nonlinear magnon dynamics, and the operation of magnonic devices~\cite{Rezende2020, Pirro2021, Wu2010, Hamadeh2014}. It is also central to quantum magnonics, where long-lived magnon excitations are essential for high cooperativity and coherent magnon--photon and magnon--phonon hybridization~\cite{Huebl2013, Yuan2022, Serha2026, Kuenstle2025}. Since these settings involve spin-wave modes with different wavelengths, propagation geometries, and mode profiles, reducing damping to a single effective Gilbert parameter~\cite{Gilbert2004} can obscure mode-dependent relaxation processes~\cite{Kim2015}. Wavenumber-dependent spin-wave relaxation has been observed in yttrium iron garnet (YIG)~\cite{GurevichAnisimov1975, Schmoll2025}, whereas data on metallic ferromagnetic films sometimes require an additional Gilbert-damping term quadratic in wavenumber~\cite{Li2016}. However, the contributions of different wavevector components to damping remain unresolved.

To address this question, we use high-resolution parametric-instability spectroscopy to selectively probe discrete thickness modes in micrometer-thick YIG films~\cite{Cherepanov1993, Arsad2023}. The measured instability threshold exhibits a regular sawtooth dependence on the applied magnetic field, reflecting successive switching of the lowest-threshold excitation channel between neighboring thickness modes. A comparison with dipole--exchange calculations shows that this structure is governed not only by the parametric coupling but, crucially, by an anisotropic wavevector dependence of spin-wave damping. We find that the damping increases systematically with the thickness-mode number, while within individual thickness branches it exhibits an approximately linear increase with the in-plane wavenumber, revealing distinct roles of the in-plane and out-of-plane wavevector components in spin-wave relaxation.

\begin{figure*}[t]
 \includegraphics[width=1\textwidth]{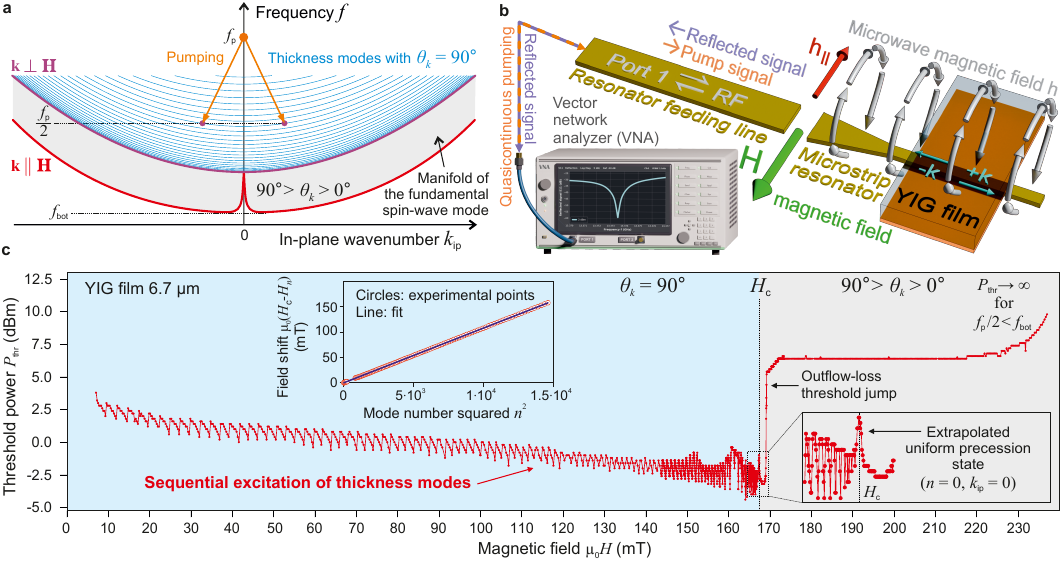}
   \caption{\label{f:setup}
Quasicontinuous parametric threshold spectroscopy of thickness-quantized magnons.
(a) Schematic spin-wave spectrum of an in-plane-magnetized YIG film under parallel pumping. Below the critical field $H_\mathrm{c}$, the resonance condition $f_k=f_\mathrm{p}/2$ is fulfilled for thickness-quantized modes with wavevectors perpendicular to the magnetization, $\theta_k = 90^\circ$ (both $k_\mathrm{oop}$ and $k_\mathrm{ip}$ are perpendicular to $\mathbf{M}_\text{s}$).
Above $H_\mathrm{c}$, the threshold is determined by the parametric excitation of states on the fundamental spin-wave branch with $90^\circ>\theta_k>0^\circ$.
(b) Experimental scheme. A microstrip half-wave resonator generates the parallel-pumping field component $h_\parallel$ in the YIG film, while a vector network analyzer measures the resonator return loss $|S_{11}|$ versus pumping power and applied field. The onset of parametric excitation increases the resonator losses and modifies the reflected signal.
(c) Representative threshold curve measured for a 6.7~\textmu m-thick YIG film at $f_\mathrm{p} \approx 13.373$\,GHz. The quasicontinuous VNA method resolves a fine sawtooth structure caused by the sequential excitation of thickness-quantized modes. The abrupt jump in the threshold above $H_\mathrm{c}$ is attributed to outflow losses from the localized pumping region. For $f_\mathrm{p}/2$ below the spin-wave-spectrum bottom $f_{\mathrm{bot}}$, no resonant spin-wave states are available, and the threshold diverges. The inset shows the field shifts $H_\mathrm{c}-H_n$ of the threshold minima (discrete dots) as a function of the squared effective mode index; the linear fit (solid line) confirms their assignment to successive thickness modes.}
\end{figure*}

The approach relies on parametric excitation, a versatile mechanism for generating and amplifying bosonic modes, with applications from quantum photonics to magnonics \cite{Asavanant2019, Serga2007, Braecher2017, Nikolaev2025}. In magnetic systems, it enables the controlled creation of magnon pairs and provides access to driven and spontaneously populated many-body states \cite{Demokritov2006, Makiuchi2024, Hioki2026, Koster2026}. In the parallel-pumping configuration~\cite{Schloemann1962} considered here, the microwave magnetic field component $h_\parallel$ parallel to the saturated static magnetization $\mathbf{M}_\mathrm{s}$ periodically modulates the effective magnetic field at the pumping frequency $f_\mathrm{p}$ and parametrically excites pairs of magnons with opposite wavevectors $\pm \mathbf{k}$ and frequency $f_\mathrm{p}/2$ (see Fig.\,\ref{f:setup}\,(a)). If the pumping frequency is fixed, sweeping the applied magnetic field shifts the spin-wave spectrum relative to $f_\mathrm{p}/2$, thereby selecting which part of the dispersion is probed and providing access to modes with different wavevectors $\mathbf{k}$. A key observable is the threshold of parametric instability, i.e., the lowest microwave power at which microwave pumping compensates the losses. This threshold encodes the interplay of damping, parametric coupling efficiency, and mode competition in the underlying spin-wave spectrum. Therefore, threshold spectroscopy can serve as a sensitive probe of spin-wave dissipation \cite{Mihalceanu2018}, provided that fine variations in the instability onset can be resolved.

Experimentally, threshold measurements are commonly performed with a microwave resonator that generates the pumping field in a magnetic sample, as schematically shown in Fig.\,\ref{f:setup}\,(b). The parametric generation of magnon pairs draws energy from the pumping field, increasing the resonator's effective losses and changing its return loss. Once the parametric instability threshold is exceeded, the parametrically excited magnon population grows exponentially in time, and the resulting change in the return loss becomes measurable.

Conventional pulsed-pumping measurements detect this onset as a kink in the reflected microwave pulse \cite{Mihalceanu2018}, but near the threshold the slow buildup of the magnon population leads to a systematic overestimation of the threshold and limits the resolution of its fine structure (Appendix~\ref{app:pulsed}).

To overcome this limitation, we develop a quasicontinuous pumping method based on a vector network analyzer (VNA), which determines the threshold from subtle changes in the resonator response (Fig.\,\ref{f:setup}\,(b)). Instead of the time-dependent reflection of a pump pulse, the VNA records the resonance curve of the loaded resonator, and the threshold is determined from changes in the return loss at the resonator frequency, which is tracked during the field and power sweeps to account for field- and power-induced resonance shifts. The high resolution in pumping power and field resolves the fine structure of the threshold curve in in-plane-magnetized YIG films (Fig.\,\ref{f:setup}\,(c)). Setup, technique, and sample are described in detail in the \hyperlink{SM}{Supplemental Material}, which includes Refs.~\cite{Bunyaev2020, Kwok1999}.

The overall shape of the measured threshold curve is determined by the spin-wave spectrum and by the competition between parallel-pumping efficiency and mode-dependent losses. In an isotropic bulk ferromagnet, the precession ellipticity, and hence the pumping coupling, is maximal for $\mathbf{k} \perp \mathbf{M}_\mathrm{s}$ and vanishes for $\mathbf{k} \parallel \mathbf{M}_\mathrm{s}$ \cite{Gurevich-Melkov1996}. In a tangentially magnetized film, shape anisotropy provides non-vanishing ellipticity even for in-plane wavevectors collinear to the static-magnetization vector.
In a spatially confined pumping geometry, as in our case, propagating spin waves experience additional outflow losses as they leave the pumped region. In our experiment, the strongly elongated pumping region of the microstrip resonator acts as a geometrical mode selector, favoring spin waves propagating close to $\theta_k=90^\circ$, for which outflow is minimized.

For $H<H_\mathrm{c}$, the condition $f_k=f_\mathrm{p}/2$ is successively fulfilled by thickness-quantized spin-wave modes with the out-of-plane wavevector $k_{\mathrm{oop}}\simeq n\pi/d$, where $n$ is the thickness-mode index and $d$ is the film thickness (see Fig.\,\ref{f:setup}\,(a)). Along a given thickness branch $n$ with $\theta_k=90^\circ$, increasing the in-plane wavevector $k_{\mathrm{ip}}$ reduces the parallel-pumping efficiency (see Appendix~\ref{app:ellipticity}), contributing to the smooth rise of the threshold. At a certain field, excitation switches to the neighboring thickness mode $n+1$ with $k_{\mathrm{ip}} \simeq 0$, restoring the stronger pumping coupling of the pure standing-wave state and producing an abrupt threshold drop. The repeated competition between increasing $k_{\mathrm{ip}}$ within a branch and switching between neighboring branches therefore gives rise to the regular sawtooth modulation resolved by the quasicontinuous VNA measurements in Fig.\,\ref{f:setup}\,(c). Similar fine structures in the reflected-power signal of parallel-pumped YIG films were previously identified as signatures of thickness-mode selectivity \cite{Kalinikos1984StandingSpinWaves, Kalinikos1985Quasisurface, Wiese1994}. Here, high-resolution VNA threshold spectroscopy turns this selectivity into a quantitative probe of anisotropic wavevector-dependent spin-wave damping.

The assignment of the threshold minima to successive thickness-quantized modes can be quantitatively validated by their field positions. For a film with uniform magnetic properties across its thickness, the exchange contribution scales as $k_{\mathrm{oop}}^2$.
We label the experimentally observed minima by $j$ and define the corresponding effective mode index as $n=j+\delta n$, where $\delta n$ accounts for the fact that the first visible minimum does not necessarily correspond to the fundamental thickness mode.
At the fixed excitation frequency $f_\mathrm{p}/2$, this gives $H_\mathrm{c}-H_n=\ell(j+\delta n)^2=\ell n^2$, with $\ell=D (\pi/d)^2$, where $D$ is the exchange constant \cite{Kittel_Introduction_to_Solid_State_Physics}.
The linear dependence of $H_\mathrm{c}-H_n$ on $n^2$ in the inset of Fig.\,\ref{f:setup}\,(c) confirms the thickness-mode assignment.

Extrapolation of the threshold minima to $n = 0$ places the uniform-precession state ($n = 0$, $k_\mathrm{ip} = 0$), and thus $H_\mathrm{c}$, at the local threshold maximum in Fig.\,\ref{f:setup}\,(c). We attribute this enhanced threshold to additional electromagnetic radiation losses and to scattering or coupling of the nearly uniform precession to other spin-wave modes.

For $H>H_\mathrm{c}$, the threshold rises sharply, passes through a broad plateau, and finally diverges. This behavior is governed by outflow losses and by the transition to perpendicular pumping and is described in Appendix~\ref{app:aboveHc}.

The broader threshold anomalies superimposed on this overall field dependence are attributed to magnon--phonon hybridization with transverse and longitudinal acoustic modes.
In the following analysis of anisotropic and wavevector-dependent spin-wave damping, we focus on the fine structure of the parallel-pumping threshold.

\begin{figure}[t]
    \includegraphics[width=1\columnwidth]{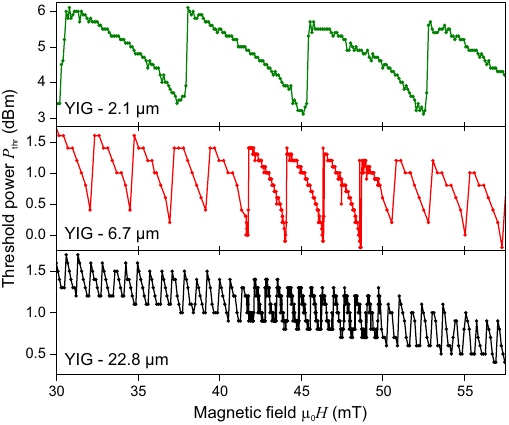}
    \caption{\label{f:2}
    Representative fragments of the parametric-instability threshold curve $P_{\mathrm{thr}}(\mu_0H)$ measured under parallel pumping for YIG films with thicknesses of 22.8, 6.7, and 2.1\,\textmu m. The traces are plotted in separate panels for clarity. The spacing between neighboring teeth increases with decreasing film thickness, as expected for modes quantized across the film thickness.
        }
\end{figure}

A key signature of the thickness-mode origin of the sawtooth is its dependence on film thickness. In Fig.\,\ref{f:2}, the tooth spacing for 22.8, 6.7, and 2.1~\textmu m-thick films measured under identical conditions increases as the film becomes thinner, as expected for modes quantized along the film normal with $k_{\mathrm{oop}} \simeq n\pi/d$.

\begin{figure}[t]
 \includegraphics[width=1\columnwidth]{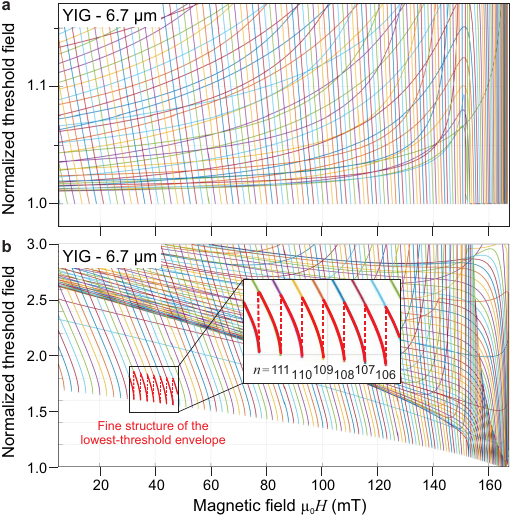}
   \caption{\label{f:theory}
Calculated parametric-instability threshold-field amplitudes for thickness-quantized spin-wave modes in a 6.7~\textmu m-thick YIG film under the experimental conditions, normalized to the absolute minimum of the threshold.
Colored curves are the thresholds of individual thickness branches; the observable threshold is the lowest of them at each field.
(a) Thresholds calculated for a mode-independent loss parameter, $\Delta H(n,k_{\mathrm{ip}})=\Delta H_0$.
The threshold minima of successive modes are then identical, so the dipole--exchange spectrum and parametric coupling alone do not produce the observed mode-index dependence.
(b) Thresholds calculated with the experimentally extracted wavevector-dependent loss parameter $\Delta H(n,k_{\mathrm{ip}})$, which depends on both the out-of-plane wavevector component $k_{\mathrm{oop}}\simeq n\pi/d$ and the in-plane wavevector component $k_{\mathrm{ip}}$.
Inset: enlarged fine structure of the lowest-threshold envelope, formed by successive switching between neighboring thickness modes.}
\end{figure}

Having established the thickness-mode assignment, we now turn to the wavevector dependence of the losses. Within parallel-pumping theory, the threshold pumping-field amplitude for a spin-wave mode $(n,k_{\mathrm{ip}})$ can be written ~\cite{Kalinikos1984StandingSpinWaves, Kostylev1995} as
\begin{equation}
h_{\mathrm{th}}(n,k_{\mathrm{ip}})=\Delta H(n,k_{\mathrm{ip}})/V(n,k_{\mathrm{ip}}),
\label{eq:threshold_general}
\end{equation}
where $\Delta H(n,k_{\mathrm{ip}})=\omega_r(n,k_{\mathrm{ip}})/\gamma$ is the magnetic-loss parameter expressed in magnetic-field units, with $\omega_r(n,k_{\mathrm{ip}})$ being the relaxation frequency of the spin-wave mode, and $V(n,k_{\mathrm{ip}})$ is the parametric-coupling coefficient.
For the pure standing spin-wave modes corresponding to the threshold minima, $k_{\mathrm{ip}}=0$ and $k_{\mathrm{oop}}=n\pi/d$. At fixed pumping frequency, the precession ellipticity, and thus the coupling coefficient for these states, is independent of the thickness-mode index, $V(n,0)=\gamma M_\mathrm{s}/\omega_\mathrm{p}$, where $M_\mathrm{s}=|\mathbf{M}_\mathrm{s}|$, $\gamma$ is the gyromagnetic ratio and $\omega_\mathrm{p}=2\pi f_\mathrm{p}$~\cite{Kalinikos1984StandingSpinWaves, Kalinikos1985Quasisurface}. Consequently, the relative variation in the threshold minima reflects the variation in losses between successive thickness modes.

This conclusion is illustrated by the calculation shown in Fig.\,\ref{f:theory}\,(a), based on the theoretical approach of Ref.\,\cite{Kostylev1995}. If the loss parameter is taken to be mode-independent, $\Delta H(n,k_{\mathrm{ip}})=\Delta H_0$, the calculated threshold minima for successive thickness modes are identical. The measured minima in Figs.~\ref{f:setup}\,(c) and~\ref{f:2}, by contrast, are clearly not constant and decrease monotonically with increasing field, i.e., with decreasing $n$. Thus, neither the dipole--exchange spectrum nor the mode dependence of the parametric coupling alone can account for this decrease. A wavevector-dependent magnon loss parameter is therefore required.

To determine whether this wavevector dependence is isotropic, we analyze the local field dependence of individual threshold teeth. Along the smooth segment of a given tooth, the thickness-mode index $n$, and hence the out-of-plane wavenumber $k_{\mathrm{oop}}$, remain fixed, whereas the in-plane wavenumber $k_{\mathrm{ip}}$ varies continuously (see Fig.\,\ref{f:2}). We convert the measured magnetic-field positions along each segment to $k_{\mathrm{ip}}$ using the calculated dipole--exchange dispersion $H_n(k_{\mathrm{ip}})$ of the corresponding thickness branch.
Since the measured finite-$k_{\mathrm{ip}}$ threshold contains contributions from both the parametric coupling and the magnetic losses, Eq.\,(1), together with $h_{\mathrm{th}}\propto\sqrt{P_{\mathrm{thr}}}$, is used to extract $\Delta H(n,k_{\mathrm{ip}})$.

\begin{figure}[b]
 \includegraphics[width=1\columnwidth]{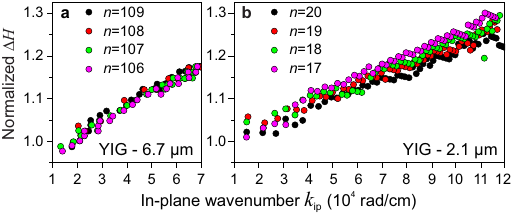}
  \caption{\label{f:theory2} Normalized magnetic-loss parameter $\Delta H$ versus in-plane wavenumber $k_{\mathrm{ip}}$ for some consecutive thickness-quantized modes in (a) the 6.7~\textmu m-thick YIG film ($n=106$--109) and (b) the 2.1~\textmu m-thick YIG film ($n=17$--20). For each thickness branch, $\Delta H$ is normalized to its value at $k_{\mathrm{ip}}=0$. The in-plane wavenumbers are obtained from the calculated dipole--exchange dispersion.}
\end{figure}

Using the experimentally extracted wavevector dependence of the magnetic-loss parameter, we then perform the full dipole--exchange threshold calculation shown in Fig.\,\ref{f:theory}\,(b).
The observable threshold is obtained as the lowest threshold among all branches at each field. The resulting lowest-threshold envelope reproduces the $n$-dependent sawtooth modulation, in contrast to Fig.\,\ref{f:theory}\,(a), demonstrating that the extracted losses consistently account for the global threshold structure.

Figure\,\ref{f:theory2} shows the extracted normalized loss for neighboring thickness modes in the 6.7~\textmu m-thick film ($n=106$--109) and the 2.1~\textmu m-thick film ($n=17$--20). In both films, the traces for neighboring modes closely follow a common dependence and exhibit an approximately linear increase with $k_{\mathrm{ip}}$. The same behavior for substantially different thicknesses and mode numbers shows that the in-plane-wavevector dependence is not specific to a particular group of modes.

Normalizing the loss parameter of each thickness branch to its value at $k_{\mathrm{ip}}=0$, the resulting dependence is well described by
\begin{equation}
\Delta H(n,k_{\mathrm{ip}})/\Delta H(n,0)=1+b_1(n)k_{\mathrm{ip}}+b_2(n)k_{\mathrm{ip}}^2.
\label{eq:normalized_loss}
\end{equation}
Here, the linear term dominates over most of the experimentally accessible range, while the quadratic contribution $b_2(n)k_{\mathrm{ip}}^2$ remains considerably smaller than $b_1(n)k_{\mathrm{ip}}$.
The finding that the leading dependence on $k_{\mathrm{ip}}$ is linear is a signature of anisotropy in $\Delta H(\mathbf{k})$.
For isotropic damping, $\Delta H$ would depend solely on the magnitude $k=|\mathbf{k}|$. For a given thickness mode,
\begin{equation}
k=\sqrt{k_{\mathrm{oop}}^2+k_{\mathrm{ip}}^2}
\simeq
k_{\mathrm{oop}}+\frac{k_{\mathrm{ip}}^2}{2k_{\mathrm{oop}}}
\qquad
(k_{\mathrm{ip}}\ll k_{\mathrm{oop}}).
\end{equation}
Accordingly, for any smooth isotropic $\Delta H(k)$, the leading term of variation with $k_{\mathrm{ip}}$ would be quadratic rather than linear.
Thus, the in-plane and out-of-plane wavevector components enter the magnetic-loss parameter differently, even though both are perpendicular to the static magnetization for the modes considered here.
The confinement across the film thickness breaks the equivalence between these two wavevector directions.

Outflow of magnons from beneath the resonator cannot cause this anisotropy, since for $H<H_\mathrm{c}$ the wavevector $k_{\mathrm{ip}}$ is directed along the resonator and the sample lies entirely within the pumped region in this direction.
A further possible contribution to the wavevector dependence of the losses is associated with changes in the precession ellipticity, which modify the relaxation rate for a fixed Gilbert damping parameter~\cite{Gurevich1973, Gurevich-Melkov1996, Verba2018, Heinz2022}. In the present case, however, this effect cannot account for the observed dependences on either $k_{\mathrm{oop}}$ or $k_{\mathrm{ip}}$ (Appendix~\ref{app:ellipticity}). 
Since confinement breaks translational symmetry across the film, a plausible source is a surface-related relaxation channel that distinguishes the in-plane and out-of-plane wavevector components. One candidate is magnon scattering at surface defects. The existing theory of this process~\cite{Nikitov1996}, however, predicts a stronger-than-linear dependence on $k_{\mathrm{ip}}$, in contrast to the approximately linear behavior observed here. This discrepancy points to another geometry-sensitive scattering channel, whose microscopic nature remains to be determined.

Beyond revealing the anisotropic wavevector dependence of losses, the present analysis also highlights which modes determine the onset of the parametric instability in films with a dense thickness-mode spectrum. In the sawtooth regime at $H<H_\mathrm{c}$, the instability is generally not governed by the fundamental spin-wave branch $n=0$. As $H$ approaches $H_\mathrm{c}$ from below, the selected states approach the uniform-precession state $(n=0,k_{\mathrm{ip}}=0)$. Away from this narrow field interval, however, the lowest threshold is set by higher-order thickness-quantized modes with $\theta_k=90^\circ$. These modes can simultaneously carry finite in-plane wavevectors; for example, in the 6.7~\textmu m-thick film, the in-plane wavenumber along the selected thickness branches reaches values exceeding $7\times10^4$\,rad/cm ($\lambda_{\text{ip}} < 900\,\mathrm{nm}$). The measured threshold curve therefore should not be interpreted as the response of a single fundamental mode, but rather as a sequence of mode-selection events between neighboring thickness branches. This conclusion applies to films with sufficiently dense thickness spectra; in submicrometer films, where the frequency separation between thickness modes is much larger, the fundamental branch may remain the lowest-threshold channel over a broad magnetic-field range \cite{Kalinikos1985Quasisurface}.

This threshold selection concerns only the onset of the instability and should be distinguished from the strongly nonlinear regime at higher pumping powers. Above threshold, nonlinear scattering processes can redistribute the initially excited magnon population over other parts of the spectrum. Wavevector-resolved BLS experiments have indeed shown that the parametrically populated spectral region broadens far beyond the dominant group selected at threshold~\cite{Serga2012}, as expected for nonlinear wave-turbulent dynamics under parametric excitation~\cite{Lvov1994}. Thus, the fundamental branch may become populated in the nonlinear regime even when it does not determine the instability threshold itself.

In summary, we have demonstrated high-resolution parametric-instability spectroscopy of thickness-quantized magnons in micrometer-thick YIG films. The observed sawtooth threshold structure arises from the sequential selection of discrete thickness modes and reflects their mode-dependent losses. Comparison with dipole--exchange threshold calculations shows that a mode-independent loss parameter would make the thresholds of the successive thickness modes identical, in clear contrast to experiment. The measured threshold modulation therefore reveals a fundamental limitation of the commonly used effective-damping description: spin-wave relaxation in magnetic films is anisotropic and wavevector dependent. The out-of-plane and in-plane wavevector components contribute differently to the losses, making anisotropic damping a key ingredient for understanding short-wavelength spin waves and nonlinear multimode magnon dynamics.

    This study was funded by the Deutsche Forschungsgemeinschaft (DFG, German Research Foundation)--TRR 173--268565370 Spin+X (Projects B01, B04 and B13), and in part by the Austrian Science Fund (FWF) [10.55776/PIN1434524]. Additional support was provided by the European Research Council (ERC) under the European Union’s Horizon Europe research and innovation programme (Consolidator Grant “MAWiCS,” Grant Agreement No. 101044526). MK acknowledges the International Collaboration Award 2026 from the University of Western Australia. OpenAI ChatGPT and Anthropic Claude were used during manuscript preparation for language editing and notation checks. All calculations, physical interpretations, and scientific conclusions were developed and verified by the authors.

\bibliography{Wavevector-dependent_damping}

@PREAMBLE{
 "\providecommand{\noopsort}[1]{}" 
 # "\providecommand{\singleletter}[1]{#1}%" 
}

@book{Rezende2020,
  author    = {Rezende, Sergio M.},
  title     = {Fundamentals of Magnonics},
  series    = {Lecture Notes in Physics},
  volume    = {969},
  publisher = {Springer},
  year      = {2020},
  doi       = {10.1007/978-3-030-41317-0}
}

@article{Pirro2021,
  title = {Advances in coherent magnonics},
  volume = {6},
  ISSN = {2058-8437},
  url = {http://dx.doi.org/10.1038/s41578-021-00332-w},
  DOI = {10.1038/s41578-021-00332-w},
  number = {12},
  journal_lng = {Nature Reviews Materials},
  journal = {Nat. Rev. Mater.},
  publisher = {Springer Science and Business Media LLC},
  author = {Pirro,  Philipp and Vasyuchka,  Vitaliy I. and Serga,  Alexander A. and Hillebrands,  Burkard},
  year = {2021},
  month = July,
  pages = {1114–1135}
}

@article{Huebl2013,
  title = {High cooperativity in coupled microwave resonator ferrimagnetic insulator hybrids},
  volume = {111},
  ISSN = {1079-7114},
  url = {http://dx.doi.org/10.1103/PhysRevLett.111.127003},
  DOI = {10.1103/physrevlett.111.127003},
  number = {12},
  journal_lng = {Physical Review Letters},
  journal = {Phys. Rev. Lett.},
  publisher = {American Physical Society (APS)},
  author = {Huebl,  Hans and Zollitsch,  Christoph W. and Lotze,  Johannes and Hocke,  Fredrik and Greifenstein,  Moritz and Marx,  Achim and Gross,  Rudolf and Goennenwein,  Sebastian T. B.},
  year = {2013},
  month = Sept,
  pages = {127003}
}

@article{Yuan2022,
  title = {Quantum magnonics: {W}hen magnon spintronics meets quantum information science},
  volume = {965},
  ISSN = {0370-1573},
  url = {http://dx.doi.org/10.1016/j.physrep.2022.03.002},
  DOI = {10.1016/j.physrep.2022.03.002},
  journal_lng = {Physics Reports},
  journal = {Phys. Rep.},
  publisher = {Elsevier BV},
  author = {Yuan,  H.Y. and Cao,  Yunshan and Kamra,  Akashdeep and Duine,  Rembert A. and Yan,  Peng},
  year = {2022},
  month = June,
  pages = {1–74}
}

@article{Serha2026,
  title = {Ultralong-living magnons in the quantum limit},
  volume = {12},
  ISSN = {2375-2548},
  url = {http://dx.doi.org/10.1126/sciadv.aee2344},
  DOI = {10.1126/sciadv.aee2344},
  number = {18},
  journal_lng = {Science Advances},
  journal = {Sci. Adv.},
  publisher = {American Association for the Advancement of Science (AAAS)},
  author = {Serha,  Rostyslav O. and McAllister,  Kaitlin H. and Majcen,  Fabian and Knauer,  Sebastian and Reimann,  Timmy and Dubs,  Carsten and Melkov,  Gennadii A. and Serga,  Alexander A. and Tyberkevych,  Vasyl S. and Chumak,  Andrii V. and Bozhko,  Dmytro A.},
  year = {2026},
  month = May,
  pages = {eaee2344}
}

@article{Gilbert2004,
  author  = {Gilbert, T. L.},
  title   = {A phenomenological theory of damping in ferromagnetic materials},
  journal_lng = {IEEE Transactions on Magnetics},
  journal = {IEEE Trans. Magn.},
  volume  = {40},
  number  = {6},
  pages   = {3443--3449},
  year    = {2004},
  doi     = {10.1109/TMAG.2004.836740}
}

@article{Asavanant2019,
  title = {Generation of time-domain-multiplexed two-dimensional cluster state},
  volume = {366},
  ISSN = {1095-9203},
  url = {http://dx.doi.org/10.1126/science.aay2645},
  DOI = {10.1126/science.aay2645},
  number = {6463},
  journal = {Science},
  publisher = {American Association for the Advancement of Science (AAAS)},
  author = {Asavanant,  Warit and Shiozawa,  Yu and Yokoyama,  Shota and Charoensombutamon,  Baramee and Emura,  Hiroki and Alexander,  Rafael N. and Takeda,  Shuntaro and Yoshikawa,  Jun-ichi and Menicucci,  Nicolas C. and Yonezawa,  Hidehiro and Furusawa,  Akira},
  year = {2019},
  month = Oct,
  pages = {373–376}
}

@article{Nikolaev2025,
  title = {Highly efficient coherent amplification of zero-field spin waves in {YIG} nanowaveguides},
  volume = {11},
  ISSN = {2375-2548},
  url = {http://dx.doi.org/10.1126/sciadv.adx2018},
  DOI = {10.1126/sciadv.adx2018},
  number = {38},
  journal_lng = {Science Advances},
  journal = {Sci. Adv.},
  publisher = {American Association for the Advancement of Science (AAAS)},
  author = {Nikolaev,  Kirill O. and Lake,  Stephanie R. and Mohapatra,  Bikash Das and Schmidt,  Georg and Demokritov,  Sergej O. and Demidov,  Vladislav E.},
  year = {2025},
  month = Sept,
  pages = {eadx2018}
}

@article{Demokritov2006,
    author = {Demokritov, S O and Demidov, V E and Dzyapko, O and Melkov, Gennadii A and Serga, A. A.  and Hillebrands, B and Slavin, Andrei N},
    DOI = {10.1038/nature05117},
    isbn = {0028-0836},
    issn = {0028-0836},
    journal_lng = {Nature},
    journal = {Nature},
    number = {7110},
    pages = {430--433},
    pmid = {17006509},
    title = {{Bose--Einstein} condensation of quasi-equilibrium magnons at room temperature under pumping},
    volume = {443},
    year = {2006}
}

@article{Serga2007,
  title = {Parametrically stimulated recovery of a microwave signal stored in standing spin-wave modes of a magnetic film},
  volume = {99},
  ISSN = {1079-7114},
  url = {http://dx.doi.org/10.1103/PhysRevLett.99.227202},
  DOI = {10.1103/physrevlett.99.227202},
  number = {22},
  journal_lng = {Physical Review Letters},
  journal = {Phys. Rev. Lett.},
  publisher = {American Physical Society (APS)},
  author = {Serga,  A. A. and Chumak,  A. V. and André,  A. and Melkov,  G. A. and Slavin,  A. N. and Demokritov,  S. O. and Hillebrands,  B.},
  year = {2007},
  month = Nov,
  pages = {227202}
}

@article{Koster2026,
  title = {Emergence of phase coherence in a magnon {B}ose--{E}instein condensate},
  volume = {22},
  ISSN = {1745-2481},
  url = {http://dx.doi.org/10.1038/s41567-026-03373-6},
  DOI = {10.1038/s41567-026-03373-6},
  number = {9},
  journal_lng = {Nature Physics},
  journal = {Nat. Phys.},  
  publisher = {Springer Science and Business Media LLC},
  author = {Koster, Malte and Schweizer, Matthias R. and Noack, Timo and Vasyuchka, Vitaliy I. and Bozhko, Dmytro A. and Hillebrands, Burkard and Weiler, Mathias and Serga,  Alexander A. and von Freymann, Georg},
  year = {2026},
  month = July,
  pages = {1430–1435}
}

@article{Serga2012,
  author  = {Serga, A. A. and Sandweg, C. W. and Vasyuchka, V. I. and Jungfleisch, M. B. and Hillebrands, B. and Kreisel, A. and Kopietz, P. and Kostylev, M. P.},
  title   = {Brillouin light scattering spectroscopy of parametrically excited dipole-exchange magnons},
  journal = {Phys. Rev. B},
  volume  = {86},
  pages   = {134403},
  year    = {2012},
  doi     = {10.1103/PhysRevB.86.134403}
}

@article{Makiuchi2024,
  title = {Persistent magnetic coherence in magnets},
  volume = {23},
  ISSN = {1476-4660},
  url = {http://dx.doi.org/10.1038/s41563-024-01798-z},
  DOI = {10.1038/s41563-024-01798-z},
  number = {5},
  journal_lng = {Nature Materials},
  journal = {Nat. Mater.},
  publisher = {Springer Science and Business Media LLC},
  author = {Makiuchi,  T. and Hioki,  T. and Shimizu,  H. and Hoshi,  K. and Elyasi,  M. and Yamamoto,  K. and Yokoi,  N. and Serga,  A. A. and Hillebrands,  B. and Bauer,  G. E. W. and Saitoh,  E.},
  year = {2024},
  month = Feb,
  pages = {627–632}
}

@article{Hioki2026,
  title = {Single- and two-mode magnon thermal squeezing},
  volume = {22},
  ISSN = {1745-2481},
  url = {http://dx.doi.org/10.1038/s41567-026-03294-4},
  DOI = {10.1038/s41567-026-03294-4},
  number = {7},
  journal_lng = {Nature Physics},
  journal = {Nat. Phys.},   
  publisher = {Springer Science and Business Media LLC},
  author = {Hioki, Tomosato and Tojo, Kaito and Elyasi, Mehrdad and Horibe, Sohei and Shimizu, Hiroki and Hoshi, Koujiro and Makiuchi, Takahiko and Bauer, Gerrit E. W. and Saitoh, Eiji},
  year = {2026},
  month = June,
  pages = {1107–1113}
}

@article{Schloemann1962,
  title = {Longitudinal susceptibility of ferromagnets in strong rf fields},
  volume = {33},
  ISSN = {1089-7550},
  url = {http://dx.doi.org/10.1063/1.1702461},
  DOI = {10.1063/1.1702461},
  number = {2},
  journal_lng = {Journal of Applied Physics},
  journal = {J. Appl. Phys.},
  publisher = {AIP Publishing},
  author = {Schl\"{o}mann,  Ernst},
  year = {1962},
  month = Feb,
  pages = {527–534}
}

@article{Braecher2017,
  title = {Parallel pumping for magnon spintronics: {A}mplification and manipulation of magnon spin currents on the micron-scale},
  volume = {699},
  ISSN = {0370-1573},
  url = {http://dx.doi.org/10.1016/j.physrep.2017.07.003},
  DOI = {10.1016/j.physrep.2017.07.003},
  journal_lng = {Physics Reports},
  journal = {Phys. Rep.},
  publisher = {Elsevier BV},
  author = {Br\"{a}cher,  T. and Pirro,  P. and Hillebrands,  B.},
  year = {2017},
  month = June,
  pages = {1–34}
}

@article{Mihalceanu2018,
  title = {Temperature-dependent relaxation of dipole-exchange magnons in yttrium iron garnet films},
  volume = {97},
  ISSN = {2469-9969},
  url = {http://dx.doi.org/10.1103/PhysRevB.97.214405},
  DOI = {10.1103/physrevb.97.214405},
  number = {21},
  journal_lng = {Physical Review B},
  journal = {Phys. Rev. B},
  publisher = {American Physical Society (APS)},
  author = {Mihalceanu,  Laura and Vasyuchka,  Vitaliy I. and Bozhko,  Dmytro A. and Langner,  Thomas and Nechiporuk,  Alexey Yu. and Romanyuk,  Vladyslav F. and Hillebrands,  Burkard and Serga,  Alexander A.},
  year = {2018},
  month = June,
  pages = {214405}
}

@article{Neumann2009,
  title = {Field-induced transition from parallel to perpendicular parametric pumping for a microstrip transducer},
  volume = {94},
  ISSN = {1077-3118},
  url = {http://dx.doi.org/10.1063/1.3130088},
  DOI = {10.1063/1.3130088},
  number = {19},
  journal_lng = {Applied Physics Letters},
  journal = {Appl. Phys. Lett.},
  publisher = {AIP Publishing},
  author = {Neumann,  T. and Serga,  A. A. and Vasyuchka,  V. I. and Hillebrands,  B.},
  year = {2009},
  month = May,
  pages = {192502}
}

@article{Kalinikos1985Quasisurface,
author  = {Kalinikos, B. A. and Kovshikov, N. G. and Kozhus', N. V.},
title   = {Parametric excitation of a series of quasisurface spin waves in thin ferromagnetic films},
journal = {Sov. Phys. Solid State},
volume  = {27},
number  = {9},
pages   = {1681--1682},
year    = {1985},
}

@article{Kalinikos1984StandingSpinWaves,
author  = {Kalinikos, B. A. and Kovshikov, N. G. and Kozhus', N. V.},
title   = {Determination of the instability threshold of standing spin waves in yttrium iron garnet films subjected to longitudinal pumping},
journal = {Sov. Phys. Solid State},
volume  = {26},
number  = {9},
pages   = {1735--1736},
year    = {1984},
}

@article{Wiese1994,
  title = {Parallel pumping fine structure at 9.4 {GHz} for in-plane magnetized yttrium iron garnet thin films},
  volume = {75},
  ISSN = {1089-7550},
  url = {http://dx.doi.org/10.1063/1.356485},
  DOI = {10.1063/1.356485},
  number = {2},
  journal_lng = {Journal of Applied Physics},
  journal = {J. Appl. Phys.},
  publisher = {AIP Publishing},
  author = {Wiese,  G. and Buxman,  L. and Kabos,  P. and Patton,  C. E.},
  year = {1994},
  month = Jan,
  pages = {1041–1046}
}

@article{Kostylev1995,
  title = {Parallel pump spin wave instability threshold in thin ferromagnetic films},
  volume = {145},
  ISSN = {0304-8853},
  url = {http://dx.doi.org/10.1016/0304-8853(94)01612-7},
  DOI = {10.1016/0304-8853(94)01612-7},
  number = {1-2},
  journal_lng = {Journal of Magnetism and Magnetic Materials},
  journal = {J. Magn. Magn. Mater.},  
  publisher = {Elsevier BV},
  author = {Kostylev,  M. P. and Kalinikos,  B. A. and D\"{o}tsch,  H.},
  year = {1995},
  month = Mar,
  pages = {93–110}
}

@article{Cherepanov1993,
     author = {Cherepanov, Vladimir and Kolokolov, Igor and L'vov, Victor S.},
     DOI = {10.1016/0370-1573(93)90107-O},
     isbn = {0370-1573},
     issn = {03701573},
     journal_lng = {Physics Reports},
     journal = {Phys. Rep.},
     number = {3},
     pages = {81--144},
     title = {{The saga of YIG: Spectra, thermodynamics, interaction and relaxation of magnons in a complex magnet}},
     volume = {229},
     year = {1993}
}

@article{Arsad2023,
     author = {Arsad, Akmal Z and Zuhdi, Ahmad Wafi Mahmood and Ibrahim, Noor Baa'yah and Hannan, Mahammad A},
     DOI = {10.3390/app13021218},
     issn = {2076-3417},
     journal_lng = {Applied Sciences},
     journal = {Appl. Sci.},
     month = {jan},
     number = {2},
     pages = {1218},
     title = {{Recent advances in yttrium iron garnet films: Methodologies, characterization, properties, applications, and bibliometric analysis for future research directions}},
     url = {https://www.mdpi.com/2076-3417/13/2/1218},
     volume = {13},
     year = {2023}
}

@book{Gurevich-Melkov1996,
    author = {Gurevich, A. G. and Melkov, G. A.},
    DOI={10.1201/9780138748487}, 
    isbn = {0849394600},
    publisher = {CRC Press, Boca Raton},
    place={Boca Raton, New York},
    title = {{Magnetization Oscillations and Waves}},
    url={https://doi.org/10.1201/9780138748487},
    year = {1996}
}

@book{Lvov1994, 
     place={Berlin, Heidelberg}, 
     title={{Wave Turbulence Under Parametric Excitation. Applications to Magnets}},
     subtitle={{Applications to Magnets}},
     DOI={10.1007/978-3-642-75295-7}, 
     publisher={Springer-Verlag},
     series = {Springer Series in Nonlinear Dynamics},
     author={L'vov, Victor S.},
     url = {https://doi.org/10.1007/978-3-642-75295-7},
     year={1994}
}

@article{Schmoll2025,
  title = {Wavenumber-dependent magnetic losses in yttrium iron garnet–gadolinium gallium garnet heterostructures at millikelvin temperatures},
  volume = {111},
  ISSN = {2469-9969},
  url = {http://dx.doi.org/10.1103/PhysRevB.111.134428},
  DOI = {10.1103/physrevb.111.134428},
  number = {13},
  journal = {Phys. Rev. B},
  publisher = {American Physical Society (APS)},
  author = {Schmoll,  David and Voronov,  Andrey A. and Serha,  Rostyslav O. and Slobodianiuk,  Denys and Levchenko,  Khrystyna O. and Abert,  Claas and Knauer,  Sebastian and Suess,  Dieter and Verba,  Roman and Chumak,  Andrii V.},
  year = {2025},
  pages = {134428},
  month = Apr 
}

@article{Bunyaev2020,
  title = {Spin-Wave relaxation by eddy currents in $\mathrm{Y_{3}Fe_{5}O_{12}/{Pt}}$ bilayers and a way to suppress it},
  volume = {14},
  ISSN = {2331-7019},
  url = {http://dx.doi.org/10.1103/PhysRevApplied.14.024094},
  DOI = {10.1103/physrevapplied.14.024094},
  number = {2},
  journal = {Phys. Rev. Appl.},
  publisher = {American Physical Society (APS)},
  author = {Bunyaev,  Sergey A. and Serha,  Rostyslav O. and Musiienko-Shmarova,  Halyna Yu. and Kreil,  Alexander J.E. and Frey,  Pascal and Bozhko,  Dmytro A. and Vasyuchka,  Vitaliy I. and Verba,  Roman V. and Kostylev,  Mikhail and Hillebrands,  Burkard and Kakazei,  Gleb N. and Serga,  Alexander A.},
  year = {2020},
  pages = {024094},
  month = Aug 
}

@article{Kwok1999,
  title = {Characterization of high-{Q} resonators for microwave filter applications},
  volume = {47},
  ISSN = {0018-9480},
  url = {http://dx.doi.org/10.1109/22.740093},
  DOI = {10.1109/22.740093},
  number = {1},
  journal_lng = {IEEE Transactions on Microwave Theory and Techniques},
  journal = {IEEE Trans. Microw. Theory Tech.},
  publisher = {Institute of Electrical and Electronics Engineers (IEEE)},
  author = {Kwok,  R. S. and Ji-Fuh Liang},
  year = {1999},
  pages = {111–114}
}

@article{Kuehn2026,
  title = {Enhancement of magnon flux toward a {B}ose--{E}instein condensate},
  volume = {113},
  ISSN = {2469-9969},
  url = {http://dx.doi.org/10.1103/nxsh-tfy8},
  DOI = {10.1103/nxsh-tfy8},
  number = {1},
  journal_lng = {Physical Review B},
  journal = {Phys. Rev. B},
  publisher = {American Physical Society (APS)},
  author = {K\"{u}hn,  Franziska and Schweizer,  Matthias R. and Azevedo,  Tamara and Vasyuchka,  Vitaliy I. and von Freymann,  Georg and L’vov,  Victor S. and Hillebrands,  Burkard and Serga,  Alexander A.},
  year = {2026},
  month = Jan,
  pages = {014409}
}

@book{Kittel_Introduction_to_Solid_State_Physics,
     author = {Kittel, Charles},
     edition = {Eighth},
     isbn = {978-0-471-62412-7},
     publisher = {Wiley},
     place={New York},
     title = {{Introduction to Solid State Physics}},
     url={https://www.wiley.com/en-us/Introduction+to+Solid+State+Physics%2C+8th+Edition-p-9780471415268},
     year = {2004}
}

@article{Li2016,
  title = {Wave-number-dependent {G}ilbert damping in metallic ferromagnets},
  volume = {116},
  ISSN = {1079-7114},
  url = {http://dx.doi.org/10.1103/PhysRevLett.116.117602},
  DOI = {10.1103/physrevlett.116.117602},
  number = {11},
  journal_lng = {Physical Review Letters},
  journal = {Phys. Rev. Lett.},
  publisher = {American Physical Society (APS)},
  author = {Li, Y. and Bailey, W. E.},
  year = {2016},
  pages = {117602},
  month = Mar 
}

@article{Kuenstle2025,
  title = {Magnon-polaron control in a surface magnetoacoustic wave resonator},
  volume = {16},
  ISSN = {2041-1723},
  url = {http://dx.doi.org/10.1038/s41467-025-66301-x},
  DOI = {10.1038/s41467-025-66301-x},
  number = {1},
  journal_lng = {Nature Communications},
  journal = {Nat. Commun.},  
  publisher = {Springer Science and Business Media LLC},
  author = {K\"{u}nstle, Kevin and Kunz, Yannik and Moussa, Tarek and Lasinger, Katharina and Yamamoto, Kei and Pirro, Philipp and Gregg, John F. and Kamra, Akashdeep and Weiler, Mathias},
  year = {2025},
  pages = {10116},
  month = Nov 
}

@article{Verba2018,
  title = {Damping of linear spin-wave modes in magnetic nanostructures: Local,  nonlocal,  and coordinate-dependent damping},
  volume = {98},
  ISSN = {2469-9969},
  url = {http://dx.doi.org/10.1103/PhysRevB.98.104408},
  DOI = {10.1103/physrevb.98.104408},
  number = {10},
  journal_lng = {Physical Review B},
  journal = {Phys. Rev. B},  
  publisher = {American Physical Society (APS)},
  author = {Verba, Roman and Tiberkevich, Vasil and Slavin, Andrei},
  year = {2018},
  pages = {104408},
  month = Sept 
}

@article{Heinz2022,
  title = {Parametric generation of spin waves in nanoscaled magnonic conduits},
  volume = {105},
  ISSN = {2469-9969},
  url = {http://dx.doi.org/10.1103/PhysRevB.105.144424},
  DOI = {10.1103/physrevb.105.144424},
  number = {14},
  journal_lng = {Physical Review B},
  journal = {Phys. Rev. B},  
  publisher = {American Physical Society (APS)},
  author = {Heinz,  Bj\"{o}rn and Mohseni,  Morteza and Lentfert,  Akira and Verba,  Roman and Schneider,  Michael and L\"{a}gel,  Bert and Levchenko,  Khrystyna and Br\"{a}cher,  Thomas and Dubs,  Carsten and Chumak,  Andrii V. and Pirro,  Philipp},
  year = {2022},
  pages = {144424},  
  month = Apr 
}

@book{Gurevich1973,
  author    = {A. G. Gurevich},
  title     = {Magnetic Resonance in Ferrites and Antiferromagnets},
  publisher = {Nauka},
  address   = {Moscow},
  year      = {1973},
  note      = {[in Russian]}
}

@article{GurevichAnisimov1975,
  author  = {A. G. Gurevich and A. N. Anisimov},
  title   = {Intrinsic spin wave relaxation processes in yttrium iron garnets},
  journal = {Sov. Phys. JETP},
  volume  = {41},
  pages   = {336--341},
  year    = {1975}
}

@article{Nikitov1996,
  title = {Reflection and scattering of spin waves by surface roughness of a ferromagnet},
  volume = {6},
  ISSN = {1361-6676},
  url = {http://dx.doi.org/10.1080/13616679609409795},
  DOI = {10.1080/13616679609409795},
  number = {1},
  journal_lng = {Waves in Random and Complex Media},
  journal = {Waves Random Media},  
  publisher = {Informa UK Limited},
  author = {Nikitov, S. A. and Gulyaev, Yu. V. and Boardman, A. D.},
  year = {1996},
  month = Jan,
  pages = {61–71}
}

@incollection{Wu2010,
  author    = {Mingzhong Wu},
  title     = {Nonlinear spin waves in magnetic film feedback rings},
  booktitle = {Solid State Physics},
  editor    = {Robert E. Camley and Robert L. Stamps},
  volume    = {62},
  pages     = {163--224},
  publisher = {Academic Press},
  address   = {Burlington},
  year      = {2010},
  doi       = {10.1016/B978-0-12-374293-3.00003-1}
}

@article{Hamadeh2014,
  author  = {A. Hamadeh and O. {d'Allivy Kelly} and C. Hahn and H. Meley and R. Bernard and A. H. Molpeceres and V. V. Naletov and M. Viret and A. Anane and V. Cros and S. O. Demokritov and J. L. Prieto and M. Mu{\~n}oz and G. de Loubens and O. Klein},
  title   = {Full control of the spin-wave damping in a magnetic insulator using spin-orbit torque},
  journal = {Phys. Rev. Lett.},
  volume  = {113},
  pages   = {197203},
  year    = {2014},
  doi     = {10.1103/PhysRevLett.113.197203}
}

@article{Kim2015,
  author  = {Joo-Von Kim},
  title   = {Role of nonlinear anisotropic damping in the magnetization dynamics of topological solitons},
  journal = {Phys. Rev. B},
  volume  = {92},
  pages   = {014418},
  year    = {2015},
  doi     = {10.1103/PhysRevB.92.014418}
}

\section*{Appendices}
\appendix
\setcounter{secnumdepth}{1}%

\vspace{-0.5em}
\section{Pulsed-pumping threshold detection}
\label{app:pulsed}
\vspace{-0.5em}
In conventional pulsed-pumping measurements, the onset of parametric excitation manifests itself as a kink in the reflected microwave pulse \cite{Mihalceanu2018}. Such measurements are well suited for observing the subsequent growth and decay of parametrically excited magnons. However, near the threshold, the growth rate is low, and the buildup time for the magnon population to produce a measurable change in the resonator response becomes long, which results in a systematic overestimation of the threshold. Accurate threshold determination may therefore require measurements for a range of pump-pulse durations followed by extrapolation to the long-pulse limit.
This procedure is time-consuming, difficult to automate, and can fail to resolve fine threshold variations associated with the spectral and damping properties of the excited magnons. \looseness=-1

\vspace{-1.5em}
\section{Precession ellipticity\\ of thickness-quantized modes}
\label{app:ellipticity}
\vspace{-0.5em}
For finite $\mathbf{k}_{\mathrm{ip}}\perp\mathbf{M}_\mathrm{s}$, the variation of the in-plane transverse dynamic magnetization component $m_{\mathrm{ip}}$ along the film generates volume magnetostatic charges and hence a demagnetizing field acting on $m_{\mathrm{ip}}$, allowing the precession ellipticity to vary with $k_{\mathrm{ip}}$. Surface magnetostatic charges at the film interfaces and volume charges produced by the thickness variation of the out-of-plane component $m_{\mathrm{oop}}$ both contribute to the demagnetizing field acting on $m_{\mathrm{oop}}$. For pure $k_{\mathrm{ip}}=0$ thickness modes, the precession ellipticity can be expressed as $|m_{\mathrm{ip}}|/|m_{\mathrm{oop}}|=[(H_n+H_{\mathrm{ex}}+M_\mathrm{s})/(H_n+H_{\mathrm{ex}})]^{1/2}$, where $H_n$ is the external resonance field of thickness mode $n$, and $H_{\mathrm{ex}}$ is the exchange field responsible for the upward shift of the thickness-mode frequency as $n$ increases. At fixed frequency, this increase in $H_{\mathrm{ex}}$ is compensated by the corresponding decrease in $H_n$, so that $H_n+H_{\mathrm{ex}}$, and hence the precession ellipticity, remains constant. Physically, increasing $k_{\mathrm{oop}}$ redistributes the magnetostatic contribution from surface toward volume charges without changing the net demagnetizing field acting on $m_{\mathrm{oop}}$.

For the pure $k_{\mathrm{ip}}=0$ thickness modes, the precession ellipticity thus remains constant at fixed frequency and therefore cannot explain the observed dependence of $\Delta H$ on $k_{\mathrm{oop}}$. For the finite-$k_{\mathrm{ip}}$ modes of Fig.~\ref{f:theory2}, a calculation using the full dipole--exchange dispersion shows that, for a constant Gilbert damping parameter, the ellipticity-related variation of the relaxation frequency is less than 0.1\% and decreases with increasing $k_{\mathrm{ip}}$ over the experimentally accessed range, opposite to the observed increase of $\Delta H$.

\vspace{-1.0em}
\section{Threshold above the critical field}
\label{app:aboveHc}
\vspace{-0.5em}
For $H>H_\mathrm{c}$, the $\theta_k=90^\circ$ states no longer intersect the half-pumping-frequency level $f_\mathrm{p}/2$, and the resonance condition is fulfilled by obliquely propagating states of the fundamental branch.
Although longitudinal states couple more strongly to parallel pumping, energy outflow from the 50\,\textmu m-wide pumping region causes substantial losses for them, initially favoring modes with $\theta_k$ closer to $90^\circ$.
With increasing field, the selected states shift toward smaller $\theta_k$, increasing the outflow losses and producing the sharp threshold rise \cite{Neumann2009}.

At higher fields, the parallel-pumping threshold exceeds that for perpendicular pumping by the microwave-field component normal to the film, producing the broad plateau \cite{Neumann2009, Kuehn2026}. With further increasing field, the perpendicular-pumping efficiency decreases, and the threshold diverges as the bottom of the spin-wave spectrum, $f_{\mathrm{bot}}$, approaches and eventually exceeds $f_\mathrm{p}/2$.

\SupplementalMaterialSetup
\begin{center}
\hypertarget{SM}{}{\large\bfseries Supplemental Material for:\\[4pt]
Anisotropic wavevector-dependent damping of thickness-quantized magnons\par}
\vspace{8pt}
Tamara~Azevedo, Rostyslav~O.~Serha, Yannik~Kunz, Matthias~R.~Schweizer, Vitaliy~I.~Vasyuchka, Mathias~Weiler, Andrii~V.~Chumak, Burkard~Hillebrands, Mikhail~Kostylev, and Alexander~A.~Serga
\end{center}
\vspace{6pt}
\twocolumngrid
\setcounter{page}{1}
\thispagestyle{titlepage}

\section{Experimental Part}
\label{sec:experimental}

\subsection{Sample and Resonator}
\label{sec:sample}

Spin waves were excited by resonator-enhanced parametric
pumping in yttrium iron garnet
(\YIG, Y$_{3}$Fe$_{5}$O$_{12}$) films of three thicknesses
(22.8\,µm, 6.7\,µm, and 2.1\,µm), all grown by liquid-phase
epitaxy on 500\,µm-thick gadolinium gallium garnet
(\GGG, Gd$_{3}$Ga$_{5}$O$_{12}$) substrates.
The comparison of the two experimental approaches was carried
out on the 6.7\,µm film, which has a width of 1.4\,mm and a
length of 15\,mm.
Its saturation magnetization at room temperature is
$M_\mathrm{s} \approx 140$\,kA/m, and its low damping,
with a linewidth of 45\,µT and a Gilbert damping parameter
$\alpha_\textbf{G} \approx 10^{-5}$, reflects the high crystalline
quality of the film, ensuring long spin-wave lifetimes and
low parametric excitation thresholds.
A sample from the same \YIG{} wafer was characterized in Ref.\,\cite{Bunyaev2020}.

A 50\,µm-wide half-wave microstrip resonator fabricated on a low-loss
dielectric substrate (RT/duroid 6010.2LM) is used to couple microwave power into
the \YIG{} film.
Its resonant frequency is
\begin{equation}
  \fr = \frac{c}{2L\sqrt{\epsilon_\mathrm{r}}},
  \label{eq:fr}
\end{equation}
where $c$ is the speed of light, $L$ the strip length, and
$\epsilon_\mathrm{r}$ the effective dielectric constant.
In our implementation $\fr \approx 13.37$\,GHz.
The quality factor of the loaded resonator is
$Q = \fr / \Delta f_\mathrm{tot}$,
where $\Delta f_\mathrm{tot}$ is the full width at half maximum
of the $|S_{11}|$ dip.
The unloaded quality factor of the microstrip resonator was
determined as 74 by the method described in\,\cite{Kwok1999}.

The \YIG{} film is placed on the resonator strip at the
position of maximum Oersted field and aligned \textit{in situ}
using the VNA while monitoring the lineshape.
For the best signal-to-noise ratio in determining the
threshold power, the sample is positioned so that the
resonator is close to critical coupling, where the slightest
change in the damping of the system has the greatest impact
on the resonance curve.

\subsection{Resonator Coupling and Losses}
\label{sec:coupling}

The coupling coefficient $\beta$ of a resonator is defined as
the ratio of external to internal loss rates:
\begin{equation}
  \beta \;=\; \frac{\Delta f_\mathrm{ext}}{\Delta f_\mathrm{in}},
  \qquad
  \Delta f_\mathrm{tot} = \Delta f_\mathrm{ext}+\Delta f_\mathrm{in}.
  \label{eq:beta}
\end{equation}
The value of $\beta$ sets the depth of the resonance and the
return-loss level at resonance\,\cite{Kwok1999}:
\begin{equation}
  |S_{11}^{\,\mathrm{res}}|
  \;=\; \left|\frac{1-\beta}{1+\beta}\right|.
  \label{eq:S11res}
\end{equation}
Three regimes arise:
(i)~$\beta < 1$ (undercoupled): internal losses dominate, $|S_{11}^{\,\mathrm{res}}| > 0$, and the resonator does not reach full absorption;
(ii)~$\beta = 1$ (critical coupling): $|S_{11}^{\,\mathrm{res}}| \to 0$, corresponding to maximum
power transfer and full absorption;
(iii)~$\beta > 1$ (overcoupled): external losses dominate, $|S_{11}^{\,\mathrm{res}}| > 0$ again, and the resonance becomes progressively shallower and broader with stronger coupling.

In the present experiment, the external losses of the loaded resonator remain essentially constant, so $\beta$ is tuned through the internal losses by the excitation of spin waves as an additional loss channel. Once the pump power reaches and exceeds the nonlinear parametric threshold, the growing magnon population dissipates energy from the resonator and thereby drives $\beta$ downward, regardless of its initial value.

\subsection{Experimental Setups}
\label{sec:setups}

\begin{figure}[]
  \centering
  \includegraphics[width=\columnwidth]{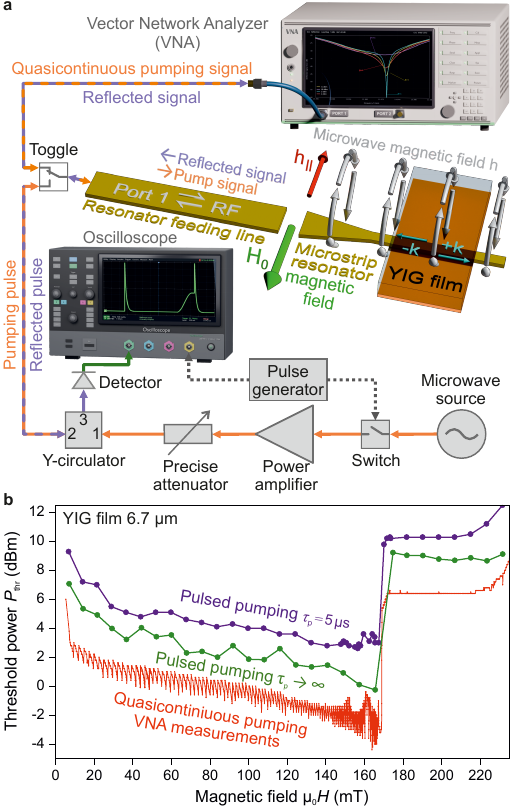}
  \caption{\label{fig:S1_setup}%
    (a)~Experimental setup for both pumping schemes. A microwave source feeds a switch and a power amplifier; a precise attenuator sets the pump power. In Setup\,1 (pulsed pumping) a pulse generator gates the switch, and the signal is routed through a Y-circulator ($1\!\to\!2$) to the stripline resonator; the reflected pulse ($2\!\to\!3$) is rectified by a diode detector and recorded on the oscilloscope. In Setup\,2 (quasicontinuous method), the VNA (Anritsu MS4647B) generates the pump and detects the reflected signal directly via Port\,1, with the path selected by the toggle switch. The \YIG{} film is placed on the stripline resonator and biased by the in-plane field $\mathbf{H}_{0}$; the Oersted-field component $h_{\parallel}$ parametrically excites magnon pairs with opposite wavevectors $\pm\mathbf{k}$. The VNA screen shows the resonance curves of the resonator depicted in \Figref{fig:S2_threshold}\,(a).
    (b)~Parametric instability threshold power $\Pthr$ as a function of the applied magnetic field $\mu_{0}H$ for the 6.7\,µm \YIG{} film, measured with the quasicontinuous VNA method (Setup\,2) and with pulsed pumping (Setup\,1) for a pulse duration $\tau_\mathrm{p}=5$\,µs and extrapolated to the continuous-wave limit $\tau_\mathrm{p}\rightarrow\infty$. The quasicontinuous method yields a lower threshold and resolves the fine structure of $\Pthr(\mu_{0}H)$ more clearly.}
\end{figure}

\subsubsection{Setup 1 -- Pulsed Pumping}

The conventional setup, used in many previous works (see, e.g., Ref.\,\cite{Mihalceanu2018}), shown in the lower half of \Figref{fig:S1_setup}\,(a), uses a microwave generator (Anritsu MG3697C, 2--67\,GHz) producing pulses of duration 5\,µs at a repetition period of 1\,ms. The signal passes through a broadband amplifier (Quinstar Technologies 12\,V) and a calibrated attenuator (HP/Agilent P382A, 14\,GHz). A Y-circulator routes the forward signal to the resonator ($1\!\to\!2$) and the reflected signal to a microwave diode detector ($2\!\to\!3$), whose DC output is read by an oscilloscope (Agilent InfiniiVision DSOX7034A). The external bias field is produced by a Lake~Shore electromagnet (power supply: Lake~Shore Model~634) and measured with a gaussmeter (Lake~Shore 475 DSP). \looseness=-1

For a given pulse, the threshold is identified as the power at which a characteristic kink appears at the trailing edge of the reflected pulse, signaling that parametric spin-wave excitation has begun to change the coupling with the resonator.
This apparent threshold depends on the pulse duration $\tau_\mathrm{p}$: because the spin-wave amplitude grows only at a finite rate above threshold, a short pulse requires a higher power for the instability to become detectable within its duration.
The duration-independent threshold corresponds to the continuous-wave limit $\tau_\mathrm{p}\rightarrow\infty$.
To estimate it with the pulsed setup, the apparent threshold is measured for a series of pulse durations and extrapolated to $\tau_\mathrm{p}\rightarrow\infty$.
This extrapolated estimate is shown as the green curve in \Figref{fig:S1_setup}\,(b), while the purple curve gives the directly measured threshold for the finite pulse duration $\tau_\mathrm{p}=5$\,µs.
As expected, the finite-duration measurement lies above the extrapolated limit at every field, since the shorter the pulse, the more the detectable instability onset is shifted to higher power.

\subsubsection{Setup 2 -- Quasicontinuous VNA Method}

In Setup\,2 (upper half of \Figref{fig:S1_setup}\,(a)), a single vector network analyzer (VNA, Anritsu MS4647B, 70\,GHz) both generates and detects the quasicontinuous microwave signal.
For each operating point, the VNA sweeps a fixed frequency window around $\fr$ and records the complex $S_{11}(f)$. The power was stepped in increments of 0.1\,dB, which sets the resolution of the extracted threshold $P_\mathrm{thr}$.
Automation is implemented in LabVIEW (program: DollRotate): the outer loop steps the bias field $\mu_{0}H$ and the inner loop steps the VNA output power, each $(H, \Pp)$ dataset being stored as a binary \textit{Technical Data Management Streaming} (.tdms) file.
A Python post-processing script then reads each file, computes $|S_{11}(f)|$, and records its minimum value $|S_{11}^{\,\mathrm{res}}|$ at resonance.
Measured example spectra are shown in \Figref{fig:S2_threshold}\,(a).

With the default intermediate-frequency bandwidth (IFBW) of 1\,kHz, the VNA measurement time is of order 1\,ms per frequency point, and in particular near the resonance frequency, two to three orders of magnitude longer than the 5\,µs pulse of Setup\,1.
The pump is therefore applied sufficiently long at each operating point to provide a close approximation to continuous excitation, allowing the threshold to be determined without pulse-duration extrapolation.
The resulting threshold is shown as the red curve in \Figref{fig:S1_setup}\,(b).

As $\mu_{0}H$ is swept, the resonance frequency $\fr$ shifts because of the field-dependent permeability of the \YIG{} film, as shown in \Figref{fig:S2_threshold}\,(b).
In the parallel geometry the measured maximum shift is $\approx 15$\,MHz.
Because the VNA sweep window is chosen wide enough to cover the full shift range, the true minimum of $|S_{11}|$ is always captured.

\subsection{Power Threshold Identification}
\label{sec:threshold}

The parametric instability threshold is the pump power $\Pp = \Pthr$ at which the spin-wave amplitude first grows exponentially, i.e., the onset of net parametric gain.
In the following we describe how $\Pthr$ is identified with the quasicontinuous VNA method (Setup\,2), where it marks the power at which effective parametric excitation increases the internal losses of the resonator and thereby changes its coupling coefficient $\beta$, producing a characteristic change in the measured $|S_{11}^{\,\mathrm{res}}|(P_\mathrm{p})$.
The precise signature depends on the initial coupling state, which is set by the position of the sample and by the bias field $\mu_{0}H$ through its control of the \YIG{} permeability and the resulting loading of the resonator (see \Figref{fig:S2_threshold}\,(a)).

\begin{figure}[]
  \centering
  \includegraphics[width=\columnwidth]{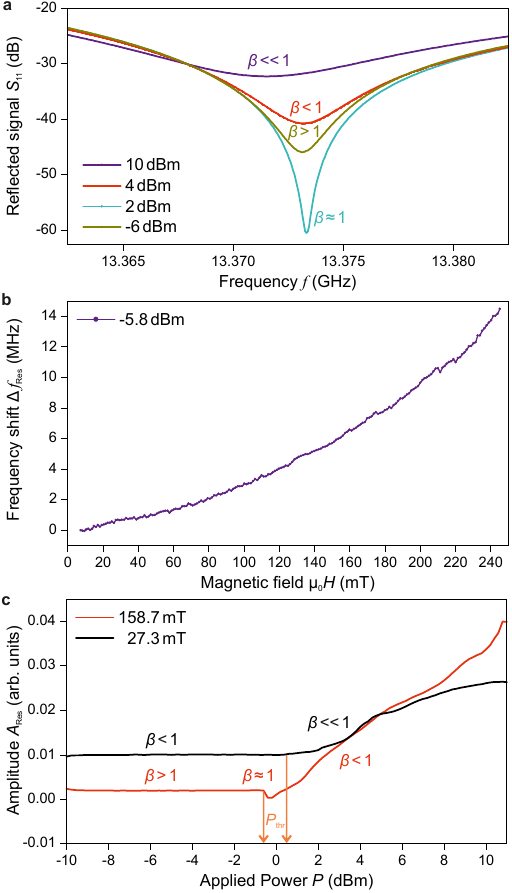}
  \caption{\label{fig:S2_threshold}%
    (a)~Reflected-signal resonance curves $|S_{11}(f)|$ of the pumping microstrip resonator, measured in dB versus the frequency $f$ at several pumping powers (-6, 2, 4, and 10\,dBm), from the subthreshold regime to well above the parametric instability threshold. As the spin-wave amplitude grows with pumping power $P_\mathrm{p}$, the internal losses increase and the coupling evolves through $\beta > 1$, $\beta \approx 1$, $\beta < 1$, and $\beta \ll 1$. The resonance first deepens toward critical coupling and then becomes shallower in the undercoupled regime.
    (b)~Subthreshold shift of the resonant frequency $\Delta f_{\mathrm{res}}$ of the pumping resonator as a function of the applied magnetic field $\mu_{0}H$, recorded at a power of -5.8\,dBm.
    (c)~Resonant amplitude $A_\mathrm{res}$ versus applied power $P$ for two external magnetic fields corresponding to different initial coupling coefficients: in the undercoupled case (27.3\,mT, $\beta < 1$), once parametric excitation sets in and the internal losses increase at the threshold power $\Pthr$, the amplitude starts to rise monotonically from the subthreshold baseline into the strongly undercoupled regime ($\beta \ll 1$); in the overcoupled case (158.7\,mT, $\beta > 1$), increasing power first drives the resonator toward critical coupling and then into the undercoupled regime. The vertical orange arrow marks the threshold power $\Pthr$, and the coupling regime at each stage is annotated.}
\end{figure}

\subsubsection{Case~A: Undercoupled initial state ($\beta < 1$)}

At lower fields the resonator can be undercoupled: internal losses exceed external losses and the $|S_{11}^{\,\mathrm{res}}|(P_\mathrm{p})$ curve stays
essentially flat as long as $P_\mathrm{p}$ remains below the threshold.
Once parametric excitation sets in at $\Pthr$, the growing spin-wave population increases the internal losses of the resonator,
so that $\beta$ drops further into the strongly undercoupled regime ($\beta \ll 1$).
This produces a kink where $|S_{11}^{\,\mathrm{res}}|$ departs from its flat subthreshold baseline and begins to change steeply (\Figref{fig:S2_threshold}\,(c), black curve), and $\Pthr$ is identified as the power at this kink.

\subsubsection{Case~B: Overcoupled initial state ($\beta > 1$)}

At high fields the resonator is initially overcoupled: external losses exceed internal losses.
Once parametric excitation sets in at $\Pthr$, the growing spin-wave population and the associated increase in internal losses drive $\beta$ downward toward critical
coupling ($\beta \approx 1$), where $|S_{11}^{\,\mathrm{res}}|$ reaches a minimum, and then beyond it into the undercoupled regime ($\beta < 1$), where $|S_{11}^{\,\mathrm{res}}|$ rises again.
Here $\Pthr$ is again identified as the lowest power at which $|S_{11}^{\,\mathrm{res}}|$ departs from the flat subthreshold baseline
(\Figref{fig:S2_threshold}\,(c), red curve).

The same coupling evolution is visible in the full lineshapes of \Figref{fig:S2_threshold}\,(a): with increasing pump power the resonance $|S_{11}|$ minimum
changes as the resonator passes through critical coupling ($\beta \approx 1$), consistent with the $\beta$ evolution described above.

\subsection{Comparison of Setup~1 and Setup~2}

Comparing $\Pthr(\mu_{0}H)$ from both setups on the same 6.7\,µm \YIG{} film (\Figref{fig:S1_setup}\,(b)), Setup~2 yields threshold powers $\approx 4$--$8$\,dBm lower across the measured field range.
The difference originates primarily from the much longer effective excitation time and the higher sensitivity to the onset of parametric excitation in the VNA measurement.
The threshold values obtained by extrapolating the pulsed measurements to $\tau_\mathrm{p}\rightarrow\infty$ also remain above the VNA values, indicating residual limitations of the finite-pulse detection and extrapolation near the instability onset.
The fine structure of the threshold curve is substantially better resolved with Setup~2 owing to its higher sensitivity and to fully automated measurements with high resolution in both pumping power and magnetic field.
A further advantage concerns the field-dependent shift of the resonance frequency: if the fixed detection frequency in Setup~1 is not recentered for each field step, this shift can move the detection point away from the optimum position on the resonance curve and raise the apparent threshold power.
Setup~2 avoids this artefact because the full $S_{11}(f)$ curve is recorded and its minimum is determined for every measurement.
In the present measurements of this work, the resonance-frequency shift accumulated over the field range of the analyzed threshold curve amounted to only about 10\,MHz (see \Figref{fig:S2_threshold}\,(b)), so that its influence on the extracted threshold was negligible.
For substantially larger shifts, however, the field dependence of $\fr$ would have to be taken into account explicitly in the data analysis.

In summary, the quasicontinuous VNA method provides substantially improved sensitivity, power and field resolution, and a high degree of automation.
This improved resolution makes it possible to resolve the fine sawtooth threshold structure analyzed in the main text.

\section{Theoretical analysis of the wavevector-dependent magnon losses}

\subsection{Parallel-pumping threshold}

To extract the magnetic-loss parameter from the measured parametric-instability threshold, we use the parallel-pumping theory developed in Ref.\,\cite{Kostylev1995}.
In this approach, the dipole--exchange spin-wave modes of an in-plane-magnetized film are characterized by the thickness-mode index $n$ and the in-plane wavenumber $k_{\mathrm{ip}}$.
The finite film thickness gives rise to the discrete out-of-plane wavenumber
$k_{\mathrm{oop}}=n\pi/d$, while $k_{\mathrm{ip}}$ varies continuously in the film plane.
The theory accounts for the dipole and exchange contributions to the spin-wave spectrum and for the mode-dependent coupling of parametrically generated magnon pairs with opposite wavevectors to the parallel pumping field.

For a given spin-wave mode $(n,k_{\mathrm{ip}})$, the threshold pumping-field amplitude can be written as
\begin{equation}
h_{\mathrm{th}}(n,k_{\mathrm{ip}})
=
\frac{\Delta H(n,k_{\mathrm{ip}})}
     {V(n,k_{\mathrm{ip}})},
\label{eq:S_threshold}
\end{equation}
where $\Delta H(n,k_{\mathrm{ip}})$ is the magnetic-loss (linewidth) parameter expressed in field units and $V(n,k_{\mathrm{ip}})$ is the parametric-coupling coefficient.
Equation\,(\ref{eq:S_threshold}) is particularly useful for the present experiment because both the spin-wave dispersion and the coupling coefficient can be calculated for every point of an experimentally resolved threshold tooth.
Thus, once a measured magnetic-field position is assigned to a particular thickness branch, the corresponding loss parameter can be obtained from the threshold using the calculated coupling coefficient.

\subsection{Losses of pure standing spin-wave modes}

The minima of the experimentally observed threshold teeth correspond to pure standing spin-wave resonance (SWR) modes with $k_{\mathrm{ip}}=0$ and $k_{\mathrm{oop}}=n\pi/d$.
Their identification is supported by the quadratic dependence of their resonance fields on the assigned thickness-mode number discussed in the main text.
For field-resolved parallel pumping at fixed pumping frequency, the coupling coefficient of these modes is independent of $n$ and is given by
\begin{equation}
V(n,0)=\frac{\gamma M_\text{s}}{\omega_\text{p}},
\label{eq:S_V_SWR}
\end{equation}
where $\gamma$ is the gyromagnetic ratio, $M_\text{s}$ is the saturation magnetization, and $\omega_p$ is the angular pumping frequency.
Consequently, the variation of the threshold amplitudes at successive tooth minima reflects the variation of the losses of the corresponding standing modes.

Choosing one standing mode $n_0$ as a reference gives
\begin{equation}
\frac{\Delta H(n,0)}{\Delta H(n_0,0)}
=
\frac{h_{\mathrm{th}}(n,0)}
     {h_{\mathrm{th}}(n_0,0)}.
\label{eq:S_normalized_SWR}
\end{equation}
For the 6.7-µm-thick film, the experimentally extracted standing-mode losses vary systematically with the thickness-mode number.
In the central range of the resolved spectrum, approximately $80\lesssim n\lesssim114$, this dependence can be represented phenomenologically by
\begin{equation}
\frac{\Delta H(n,0)}{\Delta H(n_0,0)}
=
a_0+a_1 n+a_2 n^2.
\label{eq:S_loss_n}
\end{equation}
The purpose of Eq.\,(\ref{eq:S_loss_n}) is not to assign individual microscopic relaxation mechanisms to the coefficients $a_i$, but to provide an empirical representation of the out-of-plane-wavevector dependence of the losses over the range used in the subsequent analysis.
Since $k_{\mathrm{oop}}=n\pi/d$, Eq.\,(\ref{eq:S_loss_n}) shows that the losses of the pure standing modes depend systematically on the out-of-plane wavevector component.

\subsection{Extraction of the in-plane-wavevector dependence}

The high magnetic-field resolution of the present threshold measurements allows the analysis to be extended from the discrete tooth minima to the continuous segments between them.
Each smooth segment of the sawtooth trace corresponds to a single thickness branch $n$.
Along such a segment, $n$ and therefore $k_{\mathrm{oop}}$ remain fixed, whereas the in-plane wavenumber $k_{\mathrm{ip}}$ changes continuously.
This provides a direct means of probing the in-plane-wavevector dependence of the magnetic losses without changing the thickness-mode index.

The in-plane wavenumber cannot be obtained directly from the threshold measurement.
We therefore calculate the dipole--exchange dispersion $H_n(k_{\mathrm{ip}})$ for each assigned thickness branch using the same theoretical framework as in Ref.\,\cite{Kostylev1995}.
The parameters entering the dispersion, including $\gamma$, $M_\text{s}$, and the exchange stiffness, are determined from the experimentally measured SWR resonance fields.
Numerical inversion of $H_n(k_{\mathrm{ip}})$ then converts every measured magnetic-field position along a given tooth into the corresponding $k_{\mathrm{ip}}$.

For each point $(n,k_{\mathrm{ip}})$, the same parallel-pumping calculation provides the coupling coefficient $V(n,k_{\mathrm{ip}})$.
According to Eq.\,(\ref{eq:S_threshold}), the magnetic-loss parameter is obtained as
$\Delta H(n,k_{\mathrm{ip}})=h_{\mathrm{th}}(n,k_{\mathrm{ip}})V(n,k_{\mathrm{ip}})$.
Normalizing the loss of each thickness branch to its value at the standing-wave resonance gives
\begin{equation}
\frac{\Delta H(n,k_{\mathrm{ip}})}
     {\Delta H(n,0)}
=
1+b_1(n)k_{\mathrm{ip}}+b_2(n)k_{\mathrm{ip}}^2.
\label{eq:S_loss_k}
\end{equation}
For the experimentally analyzed branches, the dependence is dominated by the term linear in $k_{\mathrm{ip}}$.
The coefficient $b_1(n)$ varies only weakly between neighboring thickness modes, whereas the contribution of the second-order term $b_2(n)$ is considerably smaller.
Thus, the approximately linear increase of the magnetic losses with $k_{\mathrm{ip}}$ is the robust feature of the local threshold analysis, while the weaker quadratic correction is mode dependent.

Figure\,\ref{f:theory2} of the main text shows the resulting normalized loss for modes $n=106-109$ in the 6.7-µm-thick film and $n=17-20$ in the 2.1-µm-thick film.
For both thicknesses, the traces of neighboring modes nearly collapse onto a common dependence and increase approximately linearly with $k_{\mathrm{ip}}$.
The persistence of this behavior for substantially different film thicknesses and thickness-mode numbers shows that the observed in-plane-wavevector dependence is not specific to a particular group of thickness modes.

\subsection{Evidence for anisotropic wavevector-dependent damping}

The dependencies obtained above allow us to test whether the magnetic losses can be described by an isotropic function of the magnitude $k=|\mathbf{k}|$ of the total spin-wave wavevector.
For the modes considered here,
\begin{equation}
k
=
\sqrt{k_{\mathrm{oop}}^2+k_{\mathrm{ip}}^2}.
\label{eq:S_ktot}
\end{equation}
If the loss parameter were isotropic in wavevector space, it could be written as $\Delta H=\Delta H(k)$, and the in-plane and out-of-plane wavevector components would not constitute independent contributions.

For $k_{\mathrm{ip}}\ll k_{\mathrm{oop}}$,
\begin{equation}
k
=
k_{\mathrm{oop}}
\sqrt{1+\frac{k_{\mathrm{ip}}^2}{k_{\mathrm{oop}}^2}}
\simeq
k_{\mathrm{oop}}
+
\frac{k_{\mathrm{ip}}^2}{2k_{\mathrm{oop}}}
+
\mathcal{O}(k_{\mathrm{ip}}^4).
\label{eq:S_ktot_expansion}
\end{equation}
Consequently, at fixed $k_{\mathrm{oop}}$, the leading variation of any smooth isotropic function $\Delta H(k)$ with $k_{\mathrm{ip}}$ must be quadratic rather than linear.
In contrast, the experimentally extracted loss parameter contains a pronounced term linear in $k_{\mathrm{ip}}$, Eq.\,(\ref{eq:S_loss_k}).
This linear contribution cannot be generated by an isotropic dependence on the magnitude of the total wavevector.

We therefore conclude that the magnetic-loss parameter depends differently on the out-of-plane and in-plane components of the spin-wave wavevector.
For the modes investigated here, both components are perpendicular to the static magnetization.
Their inequivalence is therefore not determined simply by their orientation relative to the magnetization but by the film geometry: the out-of-plane direction is confined, whereas the in-plane direction remains translationally invariant.
The observed loss thus represents anisotropic wavevector-dependent damping in the confined spin-wave system.

\subsection{Numerical reconstruction of the threshold curve}

As an independent consistency test, we calculate the complete field dependence of the parametric-instability threshold using the dipole--exchange spectrum and the corresponding parametric-coupling coefficients, computed with an advanced version of the numerical model of Ref.\,\cite{Kostylev1995}.
At a given magnetic field, several thickness branches can satisfy the parametric resonance condition, and the experimentally observed instability is determined by the branch with the lowest threshold.

First, we assume a mode- and wavevector-independent loss parameter,
$\Delta H(n,k_{\mathrm{ip}})=\Delta H_0$.
As shown in Fig.\,\ref{f:theory}\,(a) of the main text, this calculation yields identical threshold minima for the successive standing-wave resonances.
The dipole--exchange spectrum and the variation of the parametric-coupling coefficient therefore do not by themselves reproduce the experimentally observed systematic evolution of the threshold minima.

The wavevector-dependent calculation combines the experimentally determined standing-mode dependence $\Delta H(n,0)$ with the local in-plane dependence,
$\Delta H(n,k_{\mathrm{ip}})
=
\Delta H(n,0)
[1+b_1(n)k_{\mathrm{ip}}+b_2(n)k_{\mathrm{ip}}^2]$.
The thresholds of the individual thickness branches therefore become mode and wavevector dependent, and the observable threshold is obtained as their lower envelope.
The resulting calculation reproduces the characteristic $n$-dependent sawtooth modulation, as shown in Fig.\,\ref{f:theory}\,(b) of the main text.
The agreement between the locally extracted wavevector dependence and the global threshold behavior provides a consistency check that anisotropic wavevector-dependent losses are required to account for the experimentally observed fine structure.

\end{document}